\documentclass[amsmath,amssymb,twocolumn,showpacs]{revtex4}

\usepackage{bm}
\usepackage[final]{graphicx}
\usepackage{epsfig}
\usepackage[thinspace]{SIunits}
\usepackage{color}
\usepackage{amsmath}
\usepackage{ulem}

\begin{document}


\title{Origin of the resistive anisotropy in the electronic nematic phase of BaFe$_2$As$_2$ revealed by optical spectroscopy}
\date{\today}

\author{C. Mirri$^\ast$}
\affiliation{Laboratorium f\"ur Festk\"orperphysik, ETH - Z\"urich, 8093 Z\"urich, Switzerland}

\author{A. Dusza}
\affiliation{Laboratorium f\"ur Festk\"orperphysik, ETH - Z\"urich, 8093 Z\"urich, Switzerland}

\author{S. Bastelberger}
\affiliation{Laboratorium f\"ur Festk\"orperphysik, ETH - Z\"urich, 8093 Z\"urich, Switzerland}

\author{M. Chinotti}
\affiliation{Laboratorium f\"ur Festk\"orperphysik, ETH - Z\"urich, 8093 Z\"urich, Switzerland}

\author{J.-H. Chu}
\affiliation{Geballe Laboratory for Advanced Materials and Department of Applied Physics, Stanford University, Stanford CA 94305, USA}
\affiliation{Stanford Institute for Materials and Energy Sciences, SLAC National Accelerator Laboratory, 2575 Sand Hill Road, Menlo Park CA 94025, USA}

\author{H.-H. Kuo}
\affiliation{Geballe Laboratory for Advanced Materials and Department of Applied Physics, Stanford University, Stanford CA 94305, USA}
\affiliation{Stanford Institute for Materials and Energy Sciences, SLAC National Accelerator Laboratory, 2575 Sand Hill Road, Menlo Park CA 94025, USA}

\author{I.R. Fisher}
\affiliation{Geballe Laboratory for Advanced Materials and Department of Applied Physics, Stanford University, Stanford CA 94305, USA}
\affiliation{Stanford Institute for Materials and Energy Sciences, SLAC National Accelerator Laboratory, 2575 Sand Hill Road, Menlo Park CA 94025, USA}

\author{L. Degiorgi$^\ast$}
\affiliation{Laboratorium f\"ur Festk\"orperphysik, ETH - Z\"urich, 8093 Z\"urich, Switzerland}



\begin{abstract}
We perform, as a function of uniaxial stress, an optical-reflectivity investigation of the representative 'parent' ferropnictide BaFe$_2$As$_2$ in a broad spectral range, across the tetragonal-to-orthorhombic phase transition and the onset of the long-range antiferromagnetic order (AFM). The infrared response reveals that the $dc$ transport anisotropy in the orthorhombic AFM state is determined by the interplay between the Drude spectral weight and the scattering rate, but that the dominant effect is clearly associated with the metallic spectral weight. In the paramagnetic tetragonal phase, though, the $dc$ resistivity anisotropy of strained samples is almost exclusively due to stress-induced changes in the Drude weight rather than in the scattering rate, definitively establishing the anisotropy of the Fermi surface parameters as the primary effect driving the $dc$ transport properties in the electronic nematic state.
\end{abstract}

\maketitle

\newpage
Underdoped compositions of the ferropnictide superconductors exhibit a tetragonal-to-orthorhombic structural phase transition at $T_s$ that either precedes or accompanies the onset of long-range antiferromagnetic (AFM) order at $T_N$. One of the primary measurements that has lead to an understanding of the structural phase transition in terms of electronic nematic order has been the in-plane $dc$ resistivity anisotropy \cite{fisher_Science,prozorov_1,prozorov_2,fisher_ROPP}. In the orthorhombic phase this quantity suggests a substantial electronic anisotropy, while in the tetragonal phase differential elastoresistance measurements (i.e., measurements of the induced resistivity anisotropy due to anisotropic strain) reveal the diverging nematic susceptibility associated with the thermally driven nematic phase transition \cite{chu_Science,kuo}. 

There is, however, an ongoing debate as to whether the $dc$ anisotropy (both in the nematic phase ($T_N < T < T_s$) or in the tetragonal phase above $T_s$ in the presence of an external symmetry breaking field) is primarily determined by the Fermi surface (FS) or scattering rate anisotropy \cite{kuo,kuo_prl,Chuang,Allan,Nakajima_2,uchida_disorder}. Recent elastoresistivity experiments have shown that the strain-induced resistivity anisotropy in the tetragonal state of representative underdoped Fe-arsenide families is independent of disorder over a wide range of defect and impurity concentrations and consequently is not due to elastic scattering from anisotropic defects \cite{kuo,kuo_prl}. Nonetheless, measurements of annealed crystals of Ba(Fe$_{1-x}$Co$_x$)$_2$As$_2$ held under constant yet unknown uniaxial stress indicate that the resistivity anisotropy diminishes after annealing, and therefore suggest that elastic scattering might be significant in determining the resistivity anisotropy in the AFM state \cite{Nakajima_2,uchida_disorder}. Furthermore, STM measurements \cite{Chuang,Allan} at very low temperatures reveal extended anisotropic defects (i.e., nematogens), perhaps associated with impurities which locally polarize the electronic structure. Both of these observations indicate that the resistivity anisotropy might alternatively be associated with anisotropic elastic scattering from nematogens. Theoretical arguments have been made supporting this perspective \cite{gastasioro_2}, but it is far from clear how relevant the suggested mechanism is for the actual material. 

In order to clarify the microscopic origin of the resistivity anisotropy in the electronic nematic phase,
more thorough experimental studies targeting the impact
of the Fermi surface and the quasiparticle scattering rates
are desired for temperatures below $T_s$. Unfortunately, such a study is hampered by twin domain formation below $T_s$ \cite{fisher_ROPP,Tanatar_2009} and also by Fermi surface reconstruction in the AFM state. However, measurement of strain-induced optical anisotropy in the high-temperature tetragonal phase (i.e., for $T > T_s$) can directly address the same questions, circumventing both practical concerns. The microscopic mechanisms that result in the electronic anisotropy in the paramagnetic orthorhombic phase (i.e., the nematic phase) are the same as those in the strained tetragonal phase \cite{prozorov_2,kuo,kuo_prl}. In other words, measurements of the anisotropy of the optical conductivity for strained samples in the tetragonal phase directly connects to the electronic nematicity, and yet, of particular importance, is not affected by either FS reconstruction or twin boundary motion (since the material is homogeneous in the tetragonal phase).

Here, we address the controversial debate on the $dc$ anisotropy via measurements of the optical reflectivity ($R(\omega)$) of the representative 'parent' compound BaFe$_2$As$_2$, across the structural transition and upon tuning \textit{in-situ} the symmetry-breaking field represented by in-plane uniaxial compressive stress. The $R(\omega)$ measurement is an excellent probe in order to study the impact of the nematic phase in the orthorhombic state and its fluctuations in the tetragonal phase on the charge dynamics and electronic properties, since (unlike $dc$ measurements) it spans a vast energy interval, extending from the FS to energies deep into the electronic structure. From $R(\omega)$ we determine the real part $\sigma_1(\omega)$ of the optical conductivity, from which we unambiguously extract the Drude weight and scattering rate in the two crystallographic directions. These quantities can be directly related to the $dc$ transport properties. Our main result is that the $dc$ resistivity anisotropy is dominated by anisotropy in the spectral weight rather than in the scattering rate. In the long-range AFM state, both the scattering rate and spectral weight anisotropy contribute to the $dc$ resistivity anisotropy, but the spectral weight is the dominant effect. However, in the tetragonal phase the strain-induced elastoresistance anisotropy is almost completely determined by the spectral weight anisotropy. This result definitively establishes that the primary effect driving the resistivity anisotropy in the paramagnetic orthorhombic state (i.e., the electronic nematic phase) is the anisotropy of the FS parameters (i.e., anisotropy in the Fermi velocity $v_F$, effective mass $m^*$ and/or Fermi wave-vector $k_F$ under an in-plane rotation of 90 degrees). In this sense, as anticipated by previous elastoresistance measurements \cite{chu_Science,kuo,kuo_prl}, the resistivity anisotropy directly connects to a thermodynamic quantity associated with the electronic nematic order.

The BaFe$_2$As$_2$ single crystal for this study was grown using a self-flux method, as described previously \cite{chu_sample}. The structural and magnetic transition occur at $T_s \sim T_N$ = 135 K (from now on denoted as $T_{s,N}$). In order to overcome the formation of dense structural twins below $T_{s,N}$ \cite{fisher_ROPP,Tanatar_2009}, which mask the anticipated in-plane optical anisotropy, we recently developed a technique using a spring bellows to tune the uniaxial stress that is exerted on samples in order to detwin them \cite{Mirri,Mirri2}. The applied uniaxial stress is quantified by the pressure ($p$) of He-gas, flushed inside the bellows in order to control its expansion. Further details about the experimental technique and setup can be found in Refs. \onlinecite{Mirri} and \onlinecite{Mirri2} as well as in Supplemental Material \cite{SM}. The in-plane $R(\omega)$ was measured from the far infrared (FIR) to the ultraviolet (UV) at nearly normal incidence \cite{grunerbook} with the electromagnetic radiation polarized along the orthorhombic antiferromagnetic \textit{a} and ferromagnetic \textit{b} axes. Data were collected in the energy interval $\omega \sim$ 60 to 7000 cm$^{-1}$, following an initial 'zero-pressure-cool' (ZPC) \cite{Mirri}. Such a protocol is reminiscent of the zero-field-cooling procedure for magnetization measurements in a ferromagnet: from above $T_{s,N}$ we cool down the sample to the selected temperature (\textit{T}), without applying any pressure. At that \textit{T}, kept fixed during the whole experiment, we progressively increase \textit{p} in steps of 0.2 bar from 0 to a maximum pressure of 0.8 bar and measure $R(\omega)$ at each step. Then, we complete the 'pressure loop' by measuring $R(\omega)$ when releasing \textit{p} back to 0 bar. The sample was thermalized at $T >> T_s$ before performing a ZPC for the next measurement at a different temperature. These data were complemented with measurements up to 40000 cm$^{-1}$ on unstressed sample at 300 K. The optical conductivity was then extracted from the $R(\omega)$ spectrum by Kramers-Kronig (KK) transformations \cite{grunerbook}.  

\begin{figure}[!htb]
\center
\includegraphics[width=8.5cm]{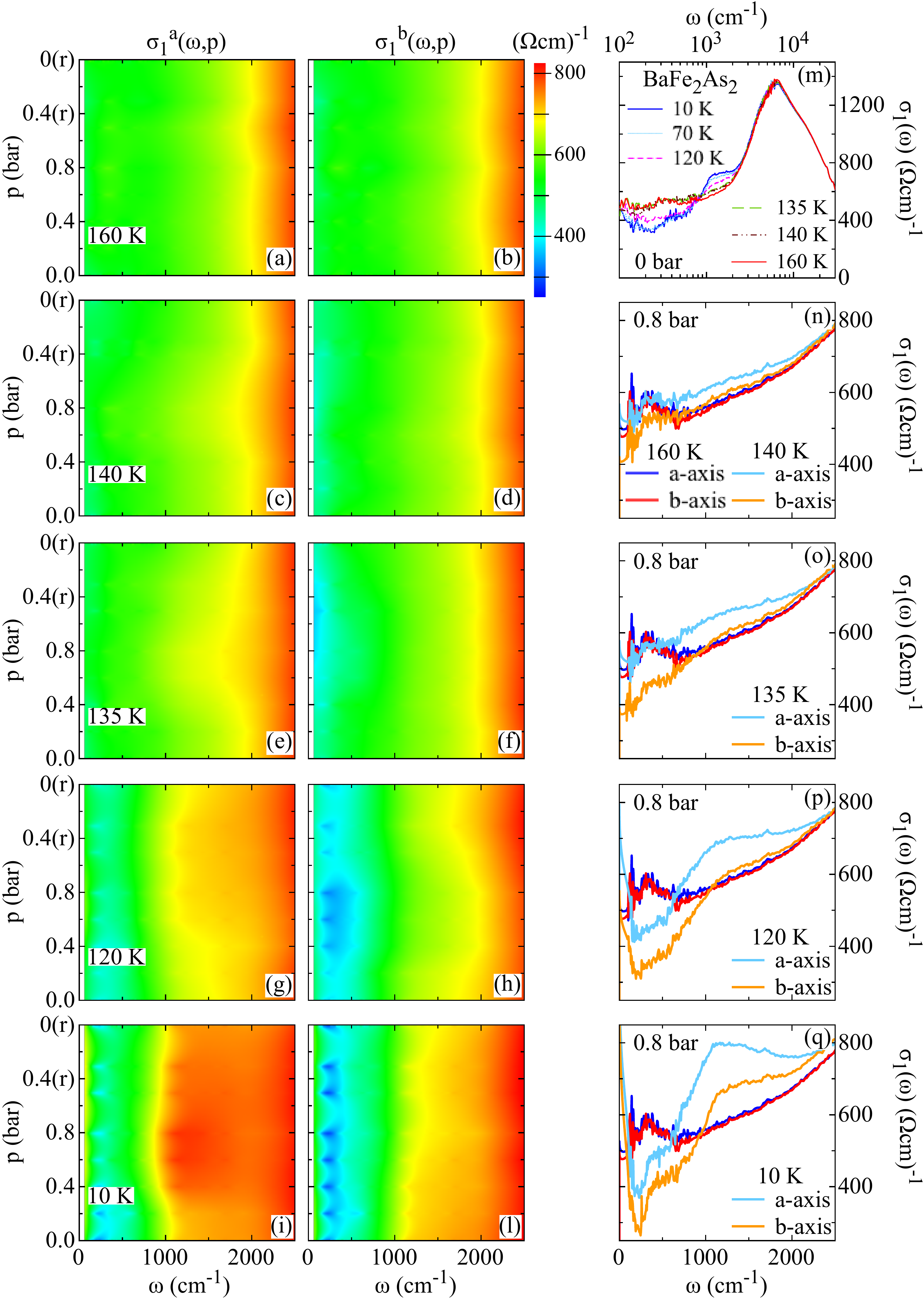}
\caption{(color online) The color maps display the pressure dependence of the real part $\sigma_1(\omega)$ of the optical conductivity up to 2500 cm$^{-1}$ along the $a$ and $b$ axis at selected temperatures: (a,b) 160 K, (c,d) 140 K, (e,f) 135 K, (g,h) 120 and (i,l) 10 K. Data have been interpolated using a first-neighbor interpolation procedure to generate the color maps. Released pressures are denoted by '(r)'. The upper right panel (m) shows the temperature dependence of $\sigma_1(\omega)$ in the FIR-UV range at 0 bar (i.e., twinned sample). The in-plane optical conductivity of BaFe$_2$As$_2$ at $p$ = 0.8 bar, compared to the data at 160 K at the same $p$, is also shown up to 2500 cm$^{-1}$ at the selected temperatures: (n) 140 K, (o) 135 K, (p) 120 K, and (q) 10 K.}
\label{fig_sigma}
\end{figure}

Figure \ref{fig_sigma} displays the real part $\sigma_1(\omega)$ of the optical conductivity. The upper right panel (m) (in a semi-log scale) emphasizes the temperature dependence of $\sigma_1(\omega)$ for the unstressed (twinned) specimen, in overall good agreement with our previous results \cite{Lucarelli_tw}. At $T \geq T_{s,N}$, $\sigma_1(\omega)$ is weakly temperature dependent, tending to a constant value in the FIR range, as is common for a conducting material. Below $T_{s,N}$, the transition into the long-range AFM state opens a partial gap, which depletes $\sigma_1(\omega)$ in FIR below 1000 cm$^{-1}$ and, due to the reshuffled spectral weight, leads to an enhancement at mid-infrared (MIR) frequencies, forming a peak centered at about 1300 cm$^{-1}$. The un-gapped portion of the FS contributes to the metallic response of $\sigma_1(\omega)$ finally merging into a narrow zero-energy mode below 200 cm$^{-1}$. The temperature dependence of $\sigma_1(\omega)$ expires above 3000 cm$^{-1}$ at the onset of the near-infrared (NIR) absorption peaked around 7000 cm$^{-1}$ and attributed to electronic interband transitions \cite{Sanna,yin}. 

The color maps of Fig. \ref{fig_sigma} reproduce the $p$ dependence of $\sigma_1(\omega)$ in the FIR-MIR spectral range ($\omega \sim$ 60 - 2500 cm$^{-1}$) at selected temperatures above, at and below $T_{s,N}$. The lower the temperature, the stronger is the optical anisotropy achieved upon applying uniaxial stress. Along the \textit{a} axis, the depletion in FIR becomes less pronounced, the intensity of the MIR absorption increases and its peak frequency shifts to lower energy upon increasing $p$. Quite the opposite behavior is observed along the \textit{b} axis (see e.g. Fig. \ref{fig_sigma}(i,l)). The resulting optical anisotropy saturates at 0.8 bar for $T << T_{s,N}$. The right panels (n-q) emphasize the $T$ dependence of the optical response for the single domain specimen (i.e., at saturation for $p$ = 0.8 bar), compared with $\sigma_1(\omega)$ at 160 K at the same $p$. The enhancement of the MIR absorption feature at about 1300 cm$^{-1}$ is clearly evident, primarily along the $a$ axis, as well as the stronger depletion of $\sigma_1(\omega)$ at $\omega <$ 1000 cm$^{-1}$ along the $b$ axis with decreasing temperature below $T_{s,N}$. Additionally, we remark the low temperature narrowing of $\sigma_1(\omega)$ at $\omega <$ 300 cm$^{-1}$ (Fig. \ref{fig_sigma}(p,q)), which is more pronounced along the $b$ axis. Upon releasing the compressive stress back to zero, a remanent anisotropy still persists at temperatures $T < T_{s,N}$ (Fig. \ref{fig_sigma}(i,l)), but it fully collapses at $T \cong T_{s,N}$ (Fig. \ref{fig_sigma}(e,f)). In other words, we can be certain that the orthorhombic state is strongly anisotropic even in the absence of any external strain. The $p$-dependent optical anisotropy at $T \le T_s$ is thus reminiscent of a hysteretic behavior \cite{Mirri,Mirri2}, which is consistent with earlier magnetoresistance measurements using an in-plane magnetic field to partially detwin single crystal samples \cite{fisher_ROPP,chu_magnetic}. Moreover, the weakly $p$-dependent stress-induced anisotropy in $\sigma_1(\omega)$ for $T > T_s$ is observable for temperatures up to at least 160 K for $p \ge$ 0.4 bar (Fig. \ref{fig_sigma}(c,d,n,e,f,o)) and is reversible upon sweeping $p$.  

\begin{figure}[!htb]
\center
\includegraphics[width=9cm]{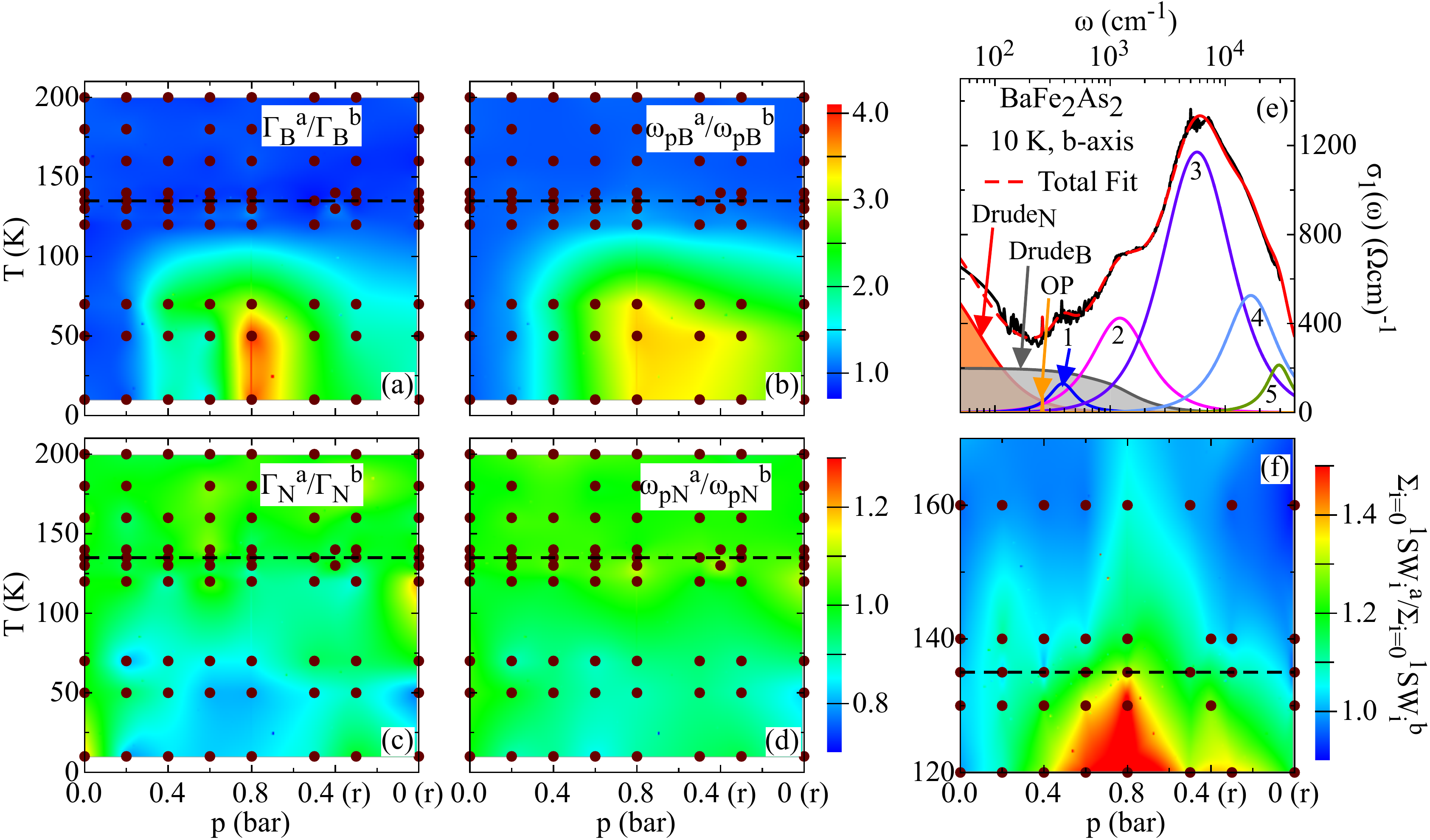}
\caption{(color online) (a-d) Pressure and temperature dependence of the anisotropy between the $a$ and $b$ axes (see text) of the plasma frequencies $\omega_{pN/B}$ and the scattering rates $\Gamma_{N/B}$ for the narrow ($N$) and broad ($B$) Drude components (Fig. S1 in Ref. \onlinecite{SM}). (e) Fit components of the optical conductivity along the \textit{b} axis in BaFe$_2$As$_2$, at 10 K and 0 bar, considered within the Drude-Lorentz approach (Eq. S1 in Ref. \onlinecite{SM}): Drude$_{N/B}$, optical phonon (OP), FIR (1), MIR (2), NIR (3), VIS-1 (4) and VIS-2 (5) h.o.'s. Apart from the OP contribution, an equivalent set of fit components is considered along the $a$ axis. (f) Pressure and temperature dependence of the anisotropy ($SW_{Drude}^a/SW_{Drude}^b$) of the total Drude weight ($SW_{Drude} = \omega_{pN}^2 + \omega_{pB}^2$, for both axes) at temperatures above 120 K. This panel, in conjunction with panels (a) and (c), reinforces the notion that the dominant effect in determining the $dc$ anisotropy induced by an external symmetry breaking field in the tetragonal phase is from changes in the FS rather than scattering. The dots indicate the fitted ($p,T$) points, which have been interpolated using a first-neighbor interpolation procedure to generate the color maps. The dashed line indicates the transition temperature $T_{s,N}$. Released pressures are denoted by '(r)'.}
\label{fig_paraDrude}
\end{figure}

In order to analyze our results, we fit the optical response functions by means of the phenomenological Drude-Lorentz model \cite{grunerbook}. The fit components for $\sigma_1(\omega)$ are displayed in Fig. \ref{fig_paraDrude}(e). The free-carrier contribution is described by a broad ($B$) and a narrow ($N$) Drude term (Drude$_B$ and Drude$_N$, respectively), mimicking the multiband nature of BaFe$_2$As$_2$. Instead of the simple Drude model that assumes a single band, the normal-state optical properties are best described by a two-Drude model that considers two separate electronic subsystems \cite{Wu}. The multiband nature of the title compound also precludes the use of the popular generalized-Drude approach commonly applied to single-band materials. At finite frequencies we add three harmonic oscillators (h.o.) FIR, MIR and NIR for the respective spectral ranges (components (1-3) in Fig. \ref{fig_paraDrude}(e)) and two high-frequency (VIS-1/2) and temperature-independent h.o.'s (components (4-5) in Fig. \ref{fig_paraDrude}(e)) covering the energy interval from the visible up to the UV range. We make use of the same set of components at all $p$ and $T$ as well as for both polarization directions, with the exception of the optical-phonon (OP) which appears only along the \textit{b} axis at $T < T_{s,N}$ \cite{Nakajima_2,schafgans}. A detailed presentation of the fit procedure and parameters as well as of its components assignment is given in Ref. \onlinecite{SM}. 

For the rest of this letter, we focus our attention on the Drude parameters (Fig. S1 in Ref. \onlinecite{SM}), which fully determine the $dc$ properties. Panels (a-d) in Fig. \ref{fig_paraDrude} display the anisotropy of the plasma frequencies ($\omega_{pN/B}$) and scattering rates ($\Gamma_{N/B}$), given by the ratio of both quantities between the two axes. The anisotropy in the scattering rates ($\Gamma_{N/B}^a/\Gamma_{N/B}^b$) only develops well below $T_{s,N}$ (Fig. \ref{fig_paraDrude}(a,c)). Indeed, above and across $T_{s,N}$ both $\Gamma_B$ and $\Gamma_N$ are fully isotropic. For $T << T_{s,N}$ $\Gamma_B$ is significantly larger (i.e., broadening of the Drude term) and $\Gamma_N$ tends to weakly decrease (i.e., narrowing of the Drude term) along the $a$ axis with respect to the $b$ axis for the fully detwinned ($p \sim$ 0.8 bar) specimen (see also Fig. S1(l,n) in Ref. \onlinecite{SM}). The anisotropy of the scattering rate $\Gamma_B$ is consistent with the well-established magnetic order (i.e., enhanced scattering with large momentum transfer along the antiferromagnetic $a$ axis) \cite{fisher_ROPP,Lucarelli}. The narrow Drude plasma frequency (Fig. \ref{fig_paraDrude}(d)) is almost fully isotropic and temperature independent (i.e., $\omega_{pN}^a/\omega_{pN}^b \sim$ 1). For the broad Drude term (Fig. \ref{fig_paraDrude}(b)), the more single-domain the sample and the lower $T$ are (see also Fig. S1(i) in Ref. \onlinecite{SM}), the larger is the plasma frequency along the $a$ axis than along the $b$ axis (i.e., $\omega_{pB}^a/\omega_{pB}^b >$ 1). Therefore, it is the broad Drude term, which mostly feels the structural and magnetic transition at $T_{s,N}$. 

We now explicitly turn our attention to the anisotropy of the total Drude weight ($SW_{Drude} = \omega_{pN}^2 + \omega_{pB}^2$), which is shown in Fig. \ref{fig_paraDrude}(f) for temperatures close to and above $T_{s,N}$. Contrary to the scattering rates, the Drude weight anisotropy (i.e., FS parameters) is strongly enhanced when approaching $T_{s,N}$ from above for finite uniaxial pressure. This FS related anisotropy grows stronger upon decreasing the temperature below $T_{s,N}$. This observation (i.e. that the dominant contribution to the anisotropy of the Drude response for $T \ge T_s$ arises from anisotropy of the spectral weight, and not anisotropy of the scattering rate) is in agreement with conclusions drawn from recent elastoresistance measurements \cite{kuo_prl} and is our main result.

\begin{figure}[!htb]
\center
\includegraphics[width=9cm]{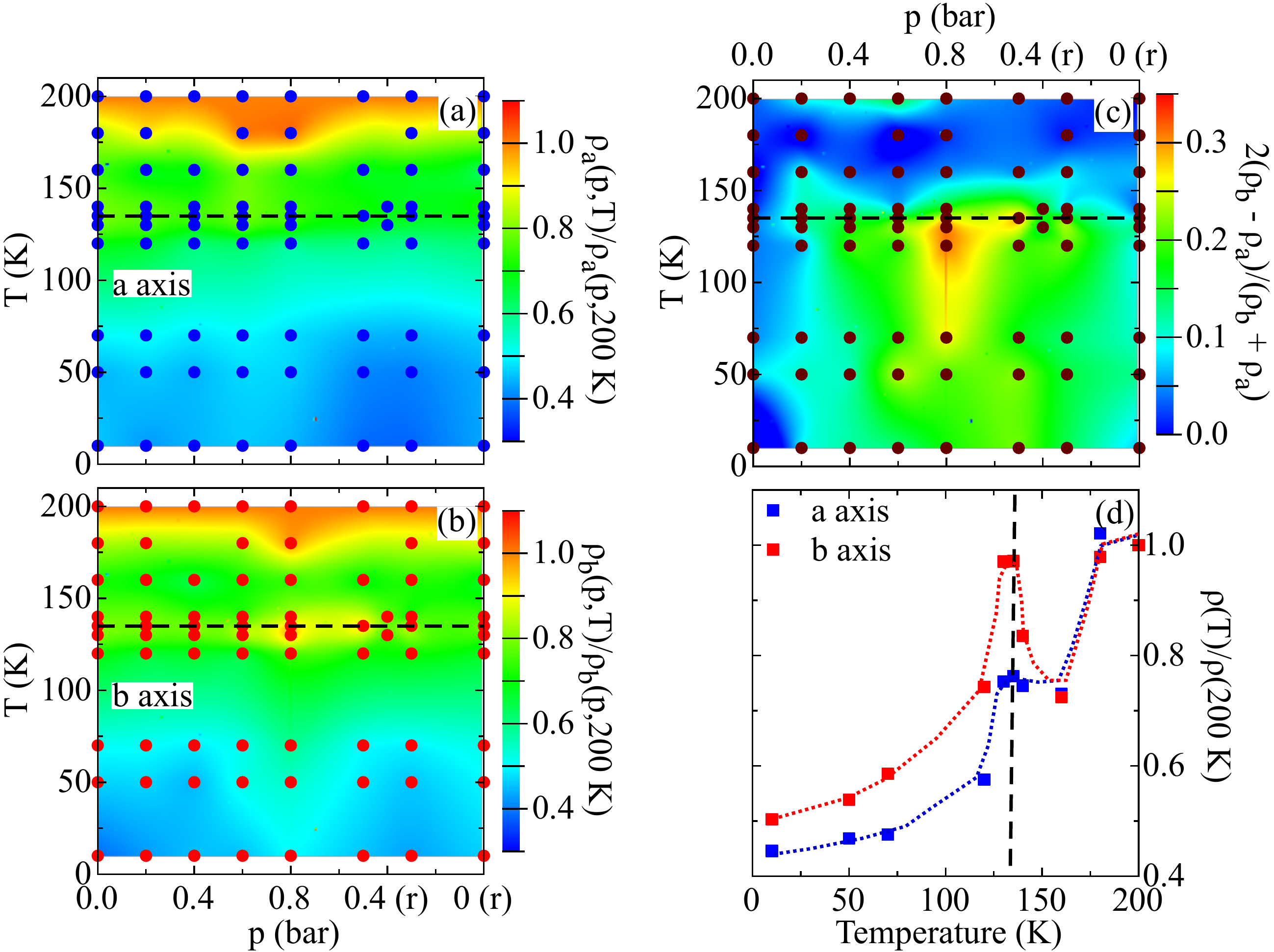}
\caption{(color online) Pressure and temperature dependence of (a) $\rho_a$ and (b) $\rho_b$ normalized at 200 K as well as (c) the $dc$ anisotropy $2(\rho_b - \rho_a)/(\rho_b + \rho_a)$, evinced from the Drude terms of the phenomenological fit (see text). The dots indicate the fitted (\textit{p},\textit{T}) points, which have been interpolated using a first-neighbor interpolation procedure to generate the color maps. Released pressures are denoted by '(r)'. (d) The temperature dependence of $\rho(T) = [\sigma_1(\omega = 0, T)]^{-1}$ from the fit of the optical conductivity at 0.8 bar (i.e., at saturation). Each resistivity curve has been normalized at 200 K. The dotted lines in (d) are guide to the eyes. The dashed lines in all panels indicate the transition temperature $T_{s,N}$.}
\label{rho_dc}
\end{figure}

We continue our analysis with the calculation of the $dc$ transport properties by reconstructing the \textit{dc} limit of the optical conductivity $\sigma_1(\omega = 0, T) = \frac{\omega_{pB}^2}{4\pi \Gamma_B} + \frac{\omega_{pN}^2}{4\pi \Gamma_N}$ from the fit parameters of the two Drude terms. Panels (a-c) of Fig. \ref{rho_dc} show the $p$ and $T$ dependence of $\rho_a$ and $\rho_b$ normalized at 200 K as well as the $dc$ anisotropy $2(\rho_b - \rho_a)/(\rho_b + \rho_a)$. Figure \ref{rho_dc}(d) displays $\rho(T) = [\sigma_1(\omega = 0, T)]^{-1}$ from the fit of the optical conductivity at 0.8 bar, which is in very good agreement with the $T$ dependence of the transport data collected on samples held under a constant uniaxial compressive stress \cite{fisher_Science}. Below $T_{s,N}$, there is an overall suppression of the $dc$ resistivity along both axes, which is more pronounced along the antiferromagnetic $a$ axis upon fully detwinning the specimen (i.e., for $p$ = 0.8 bar and upon releasing $p$ to 0.5 bar). This leads to the $dc$ anisotropy (Fig. \ref{rho_dc}(c)) which is particularly enhanced at 70 K $< T \le T_{s,N}$ for the fully detwinned sample, as observed experimentally \cite{fisher_Science}. The anisotropy of the $dc$ transport properties can also be clearly reproduced for strained samples in the tetragonal regime, for $T > T_{s,N}$. Above 0.2 bar, there is a stress-induced $dc$ anisotropy for $T_{s,N} \le T \le$ 160 K (Fig. \ref{rho_dc}(c)), which was first recognized in the transport measurement and which arises from the growing nematic susceptibility associated with the nematic phase transition \cite{chu_Science}. 

Even though the multiband nature of the iron-pnictides hampers precisely tracking the behavior of each single band crossing the Fermi level, our findings reveal a large anisotropy of the resulting conduction bands upon detwinning the specimen. This is reflected in the enhanced metallic (Drude) spectral weight along the $a$ axis with respect to the $b$ axis, which progressively evolves with external stress upon lowering $T$ (Fig. \ref{fig_paraDrude}(f)). From an experimental point of view,
it then turns out that the interplay of both plasma frequency and scattering
rate is essential for $T < T_{s,N}$, where the FS is reconstructed, when accounting for the
transport properties (Fig. \ref{rho_dc}). Significantly, though, the anisotropy in scattering rate (Fig. \ref{fig_paraDrude}(a,c)) on its own would yield a resistivity anisotropy that is opposite to that which is observed, and we therefore conclude that the dominant effect determining the transport anisotropy is in fact associated with changes in the FS parameters (Fig. \ref{fig_paraDrude}(b,d,f)).
Similarly, above the structural transition the anisotropy of the Drude weight of stressed samples (Fig. \ref{fig_paraDrude}(f)) is also primarily responsible for the anisotropic low-frequency optical and $dc$ properties.
 
Our results put some constraints on future theoretical approaches aimed reproducing
the electrodynamic response and its relationship to the $dc$ transport properties of iron-pnictides
with respect to their nematic state. Theoretical work has mostly focussed on either
anisotropic metallic scattering with spin fluctuations \cite{prelovsek,fernandes,breitkreiz,gastasioro_1,gastasioro_2,hirschfeld} or an anisotropic Drude weight
 \cite{valenzuela,liang}. It remains to be seen how our experimental results and their analysis may be
reconciled within theoretical approaches jointly addressing both Drude weight and scattering
processes. Even though impurity scattering can affect the transport anisotropy in subtle
ways, it seems to have at best only limited relevance in the electronic nematic phase. Furthermore, there is
an apparent discrepancy concerning the anisotropy of the experimental Drude scattering
rates and the conclusions drawn from recent theoretical treatments of nematogen scattering \cite{gastasioro_1}, which yield opposite trends, and hence appears to not be the dominant effect in shaping the resistivity anisotropy. In order to solve and clarify that discrepancy, a central task would be to figure out
which scattering mechanisms dominate in optical experiments. This could indicate possible
theoretical avenues in order to shed light on the interplay between scattering rate and Drude
weight, as shown by our work.

\section*{Acknowledgements}
The authors wish to thank E. Bascones, R. Fernandes, A. Chubukov, P. Hirschfeld, W. Ku, E.W. Carlson, M. Sigrist, S. Kivelson, M. Dressel, D.N. Basov and D. Lu for fruitful discussions. This work was supported by the Swiss National Science Foundation (SNSF). Work at Stanford University was supported by the
Department of Energy, Office of Basic Energy Sciences under
contract DE-AC02-76SF00515.  L.D. acknowledges the hospitality at KITP (UC Santa Barbara) within the IRONIC14 Workshop, where part of this paper was conceived. \\

$^\ast$ Correspondence and requests for materials should be addressed to: 
L. Degiorgi and C. Mirri, Laboratorium f\"ur Festk\"orperphysik, ETH - Z\"urich, 8093 Z\"urich, Switzerland; 
email: degiorgi@solid.phys.ethz.ch, chiara@phys.ethz.ch


\newcommand{\initAnhang}{
    \renewcommand{\thepage}{S\arabic{page}}
    \renewcommand{\thefigure}{S\arabic{figure}}
    \renewcommand{\thetable}{S\arabic{table}}
    \renewcommand{\theequation}{S\arabic{equation}}
    \newpage
}

\newcommand{\anhang}
{
    \setcounter{page}{1}
    \setcounter{figure}{0}
    \setcounter{table}{0}
    \newpage
}

\initAnhang
\anhang

\section*{Supplemental Material}

\subsection{Sample and experimental technique}
The as-grown BaFe$_2$As$_2$ single crystal has a square-plate shape with a thickness of 0.15 mm and a side of approximately 2 mm, with the $c$ axis perpendicular to the plane of the plate and the tetragonal $a$ axis oriented at 45$^0$ with respect to the edges of the sample, so that below $T_{s,N}$ the orthorhombic $a/b$ axes are parallel to the sides of the square \cite{fisher_ROPP}. 

We first collected data at 300 K with a Bruker IFS48 interferometer for the mid-infrared (i.e., $\omega \sim$ 500-5000 cm$^{-1}$) and a PerkinElmer Lambda 950 spectrometer up to the ultraviolet (i.e., $\omega \sim$ 3200-40000 cm$^{-1}$) range. The specimen was then mounted into the pressure device and placed inside an Oxford SM 4000 cryostat coupled to a Bruker Vertex 80v, Fourier-transform infrared interferometer for the spectral range between 60 and 7000 cm$^{-1}$. 

The pressure device, described in Ref. \onlinecite{Mirri} and \onlinecite{Mirri2}, consists of a spring bellows connected through a capillary to a pipeline outside the cryostat. The bellows can be extended/retracted by flushing He gas into its volume or evacuating it through the pipeline, thus exerting and releasing uniaxial pressure on the lateral side of the specimen (see Fig. 1 of Ref. \onlinecite{Mirri2}). The uniaxial stress, detwinning the samples, is thus applied parallel to the orthorhombic $b$ axis, which is preferentially aligned along the direction of a compressive stress, and is here declared as He-gas pressure ($p$) inside the volume of the bellows. The effective stress felt by the sample depends on its size and thickness, so that a He-gas pressure of 0.1 bar means an effective pressure on our BaFe$_2$As$_2$ crystal of about 1.5 MPa. 

For the purpose of the Kramers-Kronig (KK) transformation we made use of the standard $R(\omega) \sim \omega^{-s}$ (with $2 < s < 4$) extrapolation at high frequencies (i.e., $\omega >$ 40000 cm$^{-1}$), while the $R(\omega \rightarrow 0)$ values were obtained by merging the measured data with the Hagen-Rubens (HR) extrapolation ($R(\omega)=1-2\sqrt{\frac{\omega}{\sigma_{dc}}}$) in the energy interval between 80 and 100 cm$^{-1}$, depending from the data quality at different combinations of $T$ and $p$ \cite{grunerbook}. The $\sigma_{dc}$ values, used for the HR-extrapolation of $R(\omega)$, are consistent with the $dc$ transport measurements \cite{fisher_Science}. 

\subsection{The Drude-Lorentz fit of the optical response}
By recalling that the complex optical conductivity relates to the complex dielectric function as $\tilde{\epsilon} = \epsilon_1 + i\epsilon_2 = \epsilon_{\infty} + 4\pi i(\sigma_1 - i\sigma_2)/\omega$, we can summarize our Drude-Lorentz fit procedure (Fig. 2(e) in the main paper) as follows \cite{grunerbook}: 

\begin{equation}
\begin{split}
\tilde{\epsilon} & = \epsilon_{\infty} - \frac{\omega^2_{pN}}{\omega^2+i\omega\Gamma_N} - \frac{\omega^2_{pB}}{\omega^2+i\omega\Gamma_B} + \\
& \frac{S^2_{FIR}}{\omega^2_{0,FIR} - \omega^2 -i\omega\gamma_{FIR}} + \frac{S^2_{MIR}}{\omega^2_{0,MIR} - \omega^2 -i\omega\gamma_{MIR}} + \\
& \frac{S^2_{NIR}}{\omega^2_{0,NIR} - \omega^2 -i\omega\gamma_{NIR}} + \sum_{j=1}^2 \frac{S^2_{j}}{\omega^2_{0,j} - \omega^2 -i\omega\gamma_{j}}\\
& (+ \frac{S^2_{OP}}{\omega^2_{0,OP} - \omega^2 -i\omega\gamma_{OP}})
\end{split}
\label{eq_drudelorentz}
\end{equation}
In Eq. \ref{eq_drudelorentz}, $\epsilon_\infty$ is the optical dielectric constant. The free-carrier contribution is described by a broad (B) and a narrow (N) Drude terms (Drude$_B$ and Drude$_N$ in Fig. 2(e), respectively). At finite frequency we first add three harmonic oscillators (h.o.) FIR, MIR and NIR for the respective spectral ranges (components (1-3) in Fig. 2(e)) and then two high-frequency (VIS-1/2) and temperature-independent h.o.'s (components (4-5) in Fig. 2(e) and $j =$ 1,2 in Eq. \ref{eq_drudelorentz}) covering the energy interval from the visible up to the UV range. The last term in brackets describes the optical phonon (OP) contribution and therefore it is only added for the analysis of the optical response along the \textit{b} axis. $\Gamma_{N/B}$ and $\omega_{p N/B}$ are respectively the width at half-maximum (scattering rate) and the plasma frequency ($\omega_p = \sqrt{\frac{4\pi e^2n}{m^*}}$) of the itinerant charge carriers, with charge \textit{e}, density \textit{n} and effective mass \textit{m*}.
The parameters for each h.o. at finite frequency are the strength ($S$), the center-peak frequency ($\omega_0$) and the width ($\gamma$).
Within this phenomenological approach we simultaneously fit both $R(\omega)$ and $\sigma_1(\omega)$, achieving a good reproduction of the optical functions, as shown for example in Fig. 2(e) for $\sigma_1(\omega)$ at 10 K and 0 bar along the \textit{b} axis.

\begin{figure}[!htb]
\center
\includegraphics[width=8.5cm]{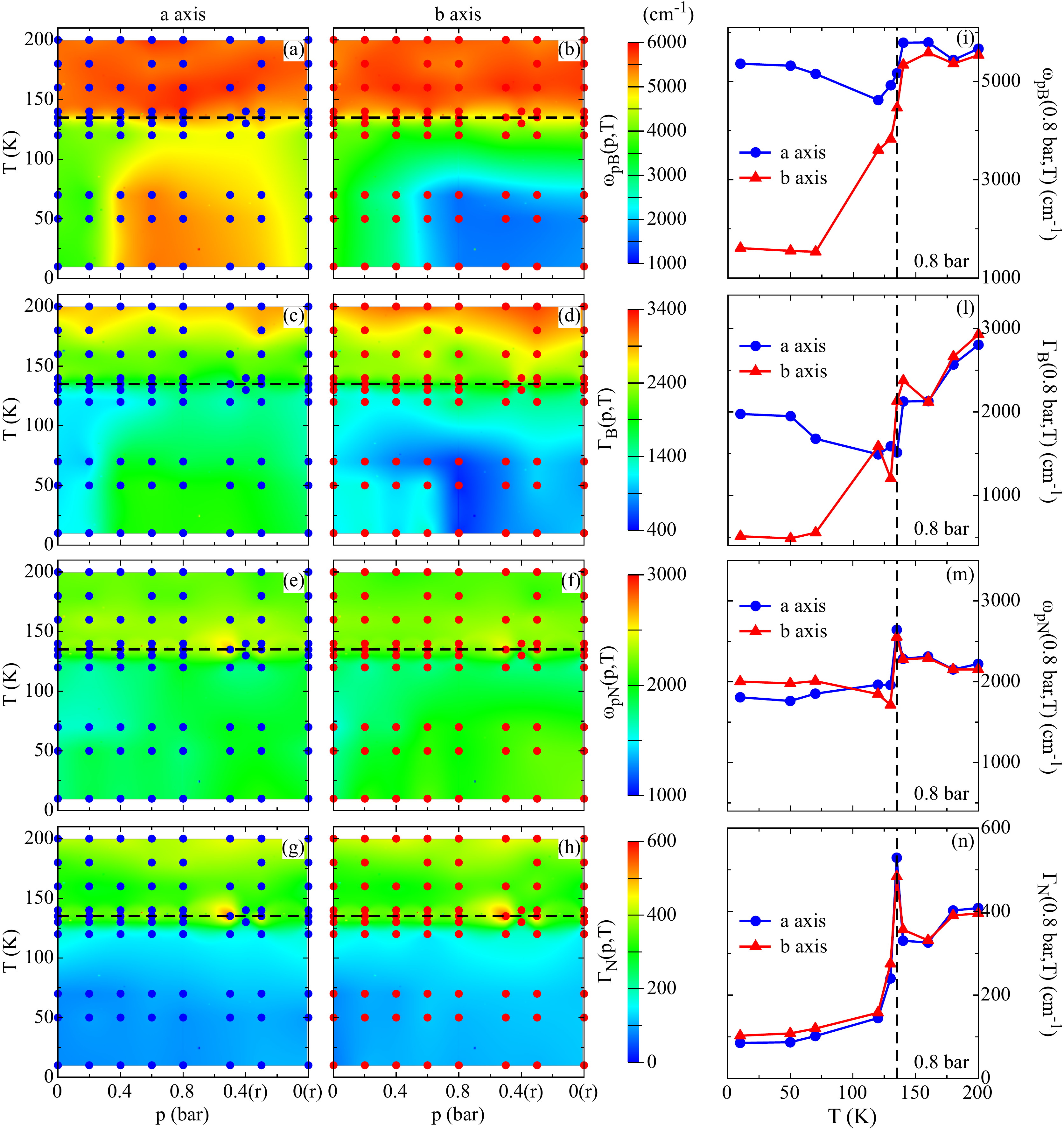}
\caption{(color online) Pressure and temperature dependence of the parameters describing the Drude components. The panels show (a, b, e, f) the plasma frequency ($\omega_{pN/B}$) and (c, d, g, h) the scattering rate ($\Gamma_{N/B}$) of the broad ($B$) and narrow ($N$) Drude term along the \textit{a} and \textit{b} axis, respectively. The dots indicate the fitted ($p,T$) points, which have been interpolated using a first-neighbor interpolation procedure to generate the color maps. Released pressures are denoted by '(r)'. Panels (i-n) display the parameters of both Drude terms at saturation (i.e., at 0.8 bar). The dashed lines in all panels indicate the transition temperature $T_{s,N}$.}
\label{fig_paramDrude}
\end{figure}

\begin{table}[!t]
\centering
\begin{tabular}{|l|c|c|c|c|c|c|c|}
\hline
\textbf{h.o.}&$\omega_{0j}$ (cm$^{-1}$)&$S_j$ (cm$^{-1}$)&$\gamma_j$ (cm$^{-1}$)\\
\hline
\textbf{VIS-1} & 17000 & 26000 & 22000 \\
\hline
\textbf{VIS-2}  & 29000 & 15000 & 18000  \\
\hline
\end{tabular}
\caption{Peak-frequency ($\omega_{0j}$), strength ($S_j$) and width ($\gamma_j$) of the two $p$- and $T$-independent h.o.'s (Eq. \ref{eq_drudelorentz}) in the visible-UV spectral range (components (4-5) in Fig. 2(e) of the main paper).} 
\label{Tab}
\end{table}

As confirmed by the optical work of several groups, it is well established \cite{Wu} that a generic low-frequency optical feature of the ferropnictide materials consists in a metallic zero energy mode, merging into a broad tail up to about 1000 cm$^{-1}$ in $\sigma_1(\omega)$ (see e.g. Fig. 1(m) in the main paper). In principle that tail can be fitted with a very broad, low-frequency h.o. This latter h.o. would have such a low resonance frequency (i.e., $\omega_0$ less than 50 cm$^{-1}$) so that it actually resembles a Drude response. While low-energy interband transitions are expected for the ferropnictide \cite{benfatto,calderon,calderon2}, they are not expected to fall below 50 cm$^{-1}$ and as such they would be rather un-physical. The multiband nature of the ferropnictide coupled with the Drude-like shape of the (artificial) low-frequency h.o. led to the pretty well established fit procedure of Eq. \ref{eq_drudelorentz} \cite{Wu}, which models the free-carriers response as two separate, uncorrelated electronic subsystems rather than a single dominant band. However, an orbital assignment of both Drude terms is not immediately obvious. The additional h.o.'s (Eq. \ref{eq_drudelorentz} and Fig. 2(e)) do then account for the electronic interband transitions, predicted on various occasions and widely discussed in the literature \cite{Sanna,yin}. Particularly, the FIR and MIR h.o.'s (components (1-2) in Fig. 2(e)) describe low energy interband transitions (anticipated above) \cite{benfatto,calderon,calderon2}, interplaying with the Drude components and finally merging into the spin-density-wave gap excitation at $T < T_{s,N}$, i.e. in the orthorhombic long-range antiferromagnetic (AFM) state \cite{Lucarelli_tw}. 

The plasma frequencies $\omega_{pN/B}$ and the scattering rates $\Gamma_{N/B}$ of the narrow and broad Drude components are shown in Fig. \ref{fig_paramDrude}(a-h) for both crystallographic directions as a function of \textit{p} and \textit{T}. Panels (i-n) of Fig. \ref{fig_paramDrude} specifically emphasize the temperature dependence of the Drude parameters at 0.8 bar (i.e., saturation). The narrow Drude component, which is obviously tight to the necessary HR extrapolation for the purpose of the KK transformations, is rather isotropic between both axes at all $T$ and for any degree of detwinning. The corresponding Drude plasma frequency $\omega_{pN}$ is $p$- and $T$-independent, while the scattering rate $\Gamma_N$ gets suppressed in a similar fashion along both axes upon crossing the structural and AFM phase transition and for any applied stress. This indicates a quite pronounced narrowing of $\sigma_1(\omega)$ below 200 cm$^{-1}$ upon lowering $T$. We may state that this narrow Drude term accounts for itinerant charge carriers, which are not gapped by the AFM transition and belong to fully isotropic band even in the orthorhombic phase. The reduced scattering below $T_{s,N}$ bears however testimony for a AFM-induced suppression of scattering channels also for that portion of Fermi surface. 

\begin{figure}[!htb]
\center
\includegraphics[width=8.5cm]{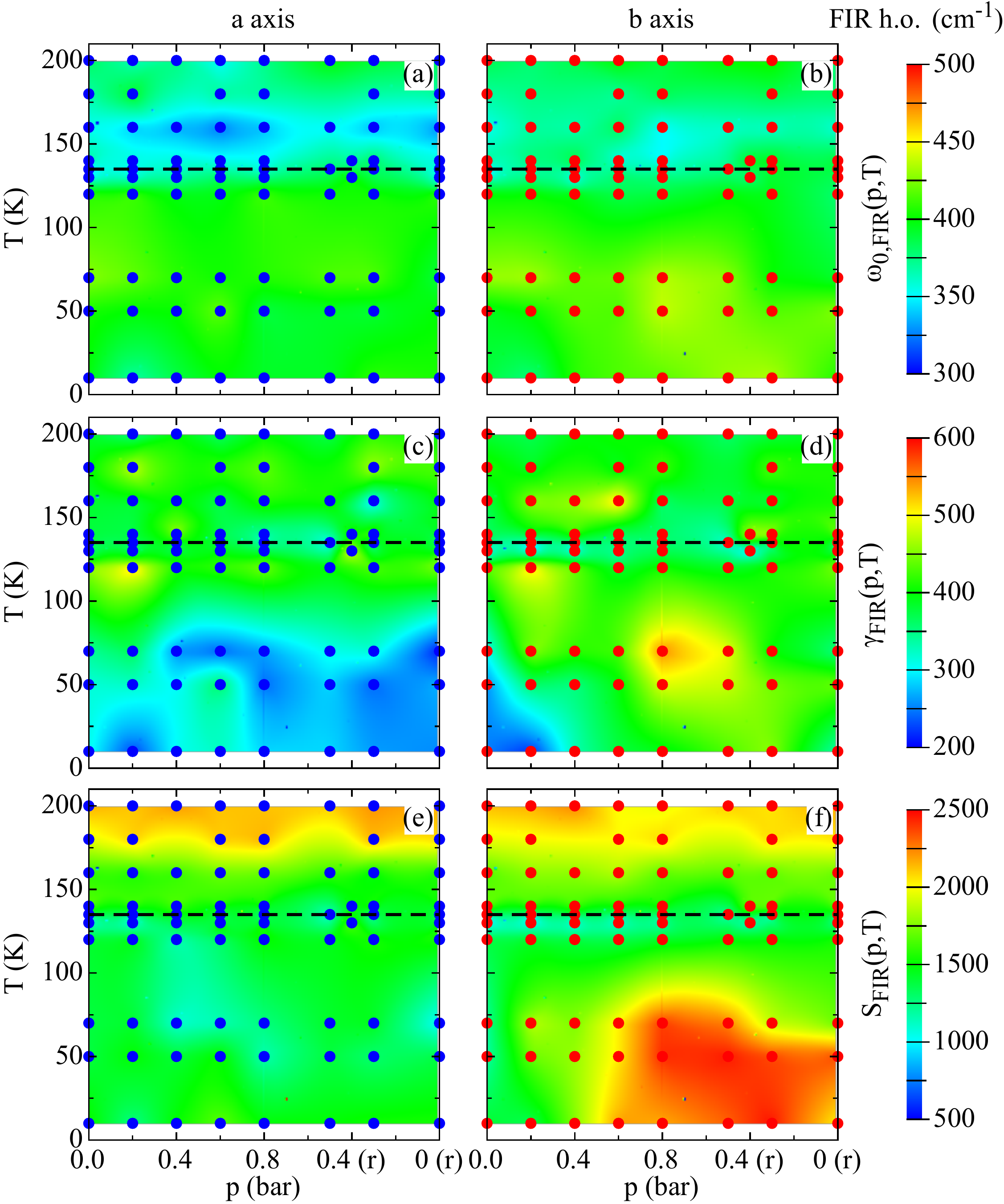}
\caption{(color online) Pressure and temperature dependence of the fit parameters for the FIR-h.o. The panels show the peak frequency $\omega_0$ (a,b), the width $\gamma$ (c,d) and the strength $S$ (e,f) for the \textit{a} and \textit{b} axis, respectively. The dots indicate the fitted (\textit{p},\textit{T}) points, which have been interpolated using a first-neighbor interpolation procedure to generate the color maps. The dashed line indicates the transition temperature $T_{s,N}$. Released pressures are denoted by '(r)'.}
\label{fig_paraFIR}
\end{figure}

On the contrary, the broad Drude term turns out to be largely anisotropic below $T_{s,N}$ and thus very much affected by the degree of detwinning. Below $T_{s,N}$ and for $p \le $ 0.2 bar, both plasma frequency and scattering rate decrease monotonically with $T$, as an effect of the Fermi surface gapping and the suppression of scattering-channels due to the AFM transition. Since the specimen is not yet fully detwinned for $p \le$ 0.2 bar, the overall depletion of both (broad) Drude parameters reflects an average-behavior of those quantities for the two crystallographic directions. Upon detwinning the specimen (i.e., for $p \ge$ 0.2 bar) the anisotropy of $\omega_{pB}$ and $\Gamma_{B}$ develops along the orthorhombic axes. Along the $a$ axis both plasma frequency and scattering rate are enhanced with respect to the $b$ axis parameters, which get further suppressed and depleted at $T < T_{s,N}$ and upon increasing $p$. We also observe that the anisotropy in $\omega_{pB}$ and $\Gamma_B$ parameters achieved at saturation (i.e., at 0.8 bar, where the sample is totally detwinned \cite{Mirri,Mirri2}) almost fully persists in the remanent phase (i.e., at zero released $p$). The result of our fits at saturation and for $T \le T_s$ is in very good agreement with the previous analysis of the effective metallic response performed on specimens held under constant uniaxial pressure \cite{Lucarelli}.

\begin{figure}[!htb]
\center
\includegraphics[width=8.5cm]{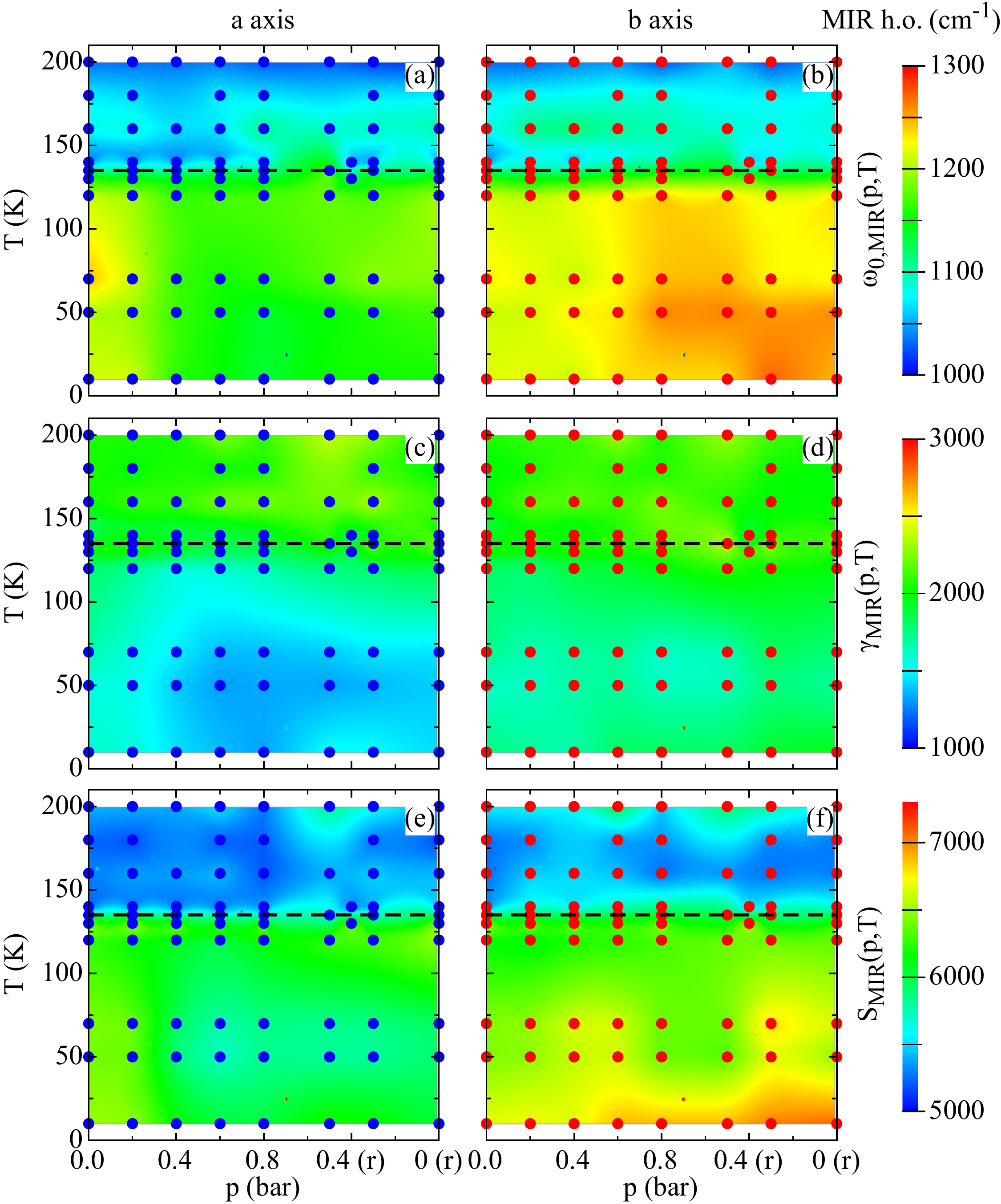}
\caption{(color online) Pressure and temperature dependence of the fit parameters for the MIR-h.o. The panels show the peak frequency $\omega_0$ (a,b), the width $\gamma$ (c,d) and the strength $S$ (e,f) for the \textit{a} and \textit{b} axis, respectively. The dots indicate the fitted (\textit{p},\textit{T}) points, which have been interpolated using a first-neighbor interpolation procedure to generate the color maps. The dashed line indicates the transition temperature $T_{s,N}$. Released pressures are denoted by '(r)'.}
\label{fig_paraMIR}
\end{figure}

\begin{figure}[!htb]
\center
\includegraphics[width=8.5cm]{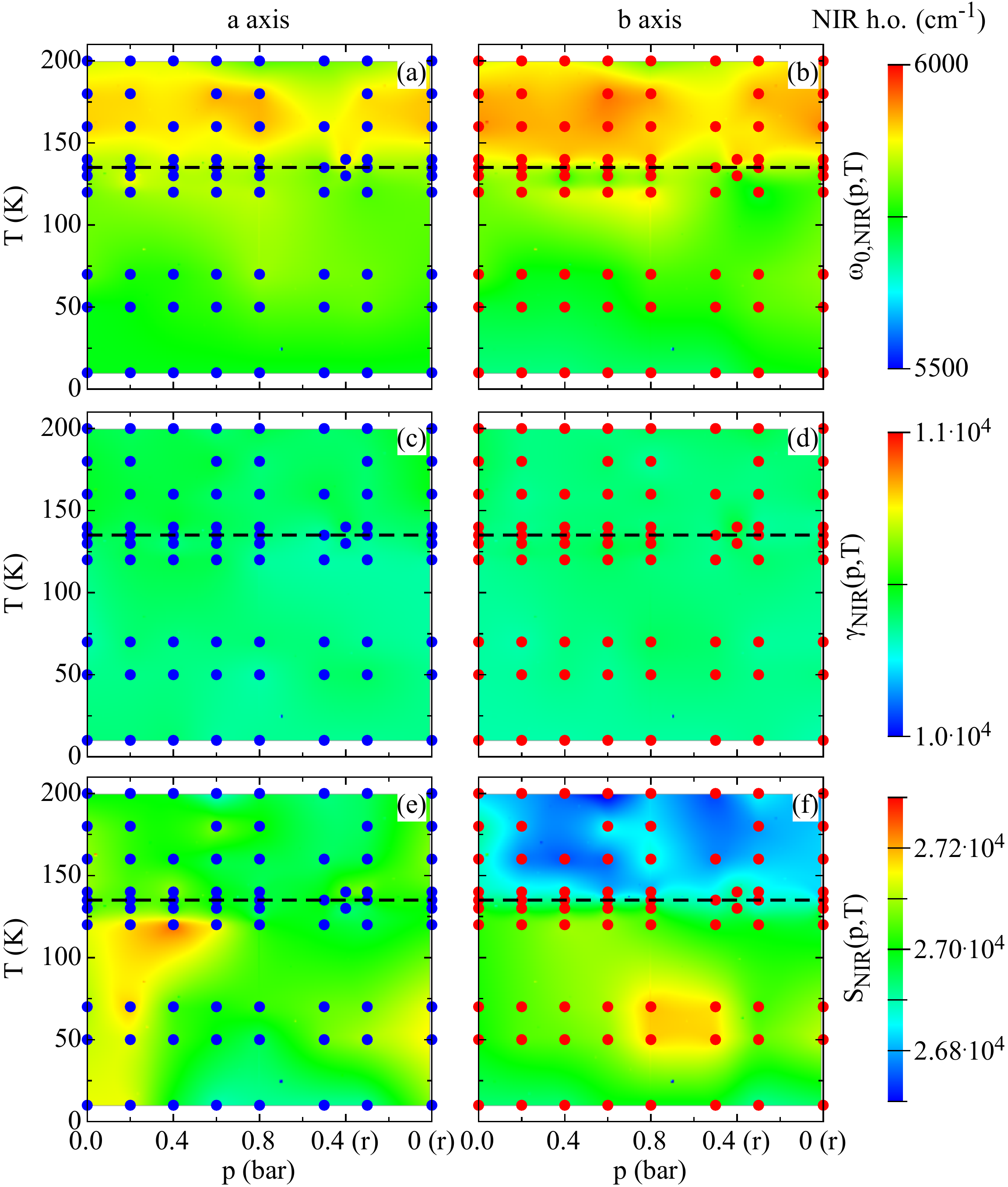}
\caption{(color online) Pressure and temperature dependence of the fit parameters for the NIR-h.o. The panels show the peak frequency $\omega_0$ (a,b), the width $\gamma$ (c,d) and the strength $S$ (e,f) for the \textit{a} and \textit{b} axis, respectively. The dots indicate the fitted (\textit{p},\textit{T}) points, which have been interpolated using a first-neighbor interpolation procedure to generate the color maps. The dashed line indicates the transition temperature $T_{s,N}$. Released pressures are denoted by '(r)'.}
\label{fig_paraNIR}
\end{figure}

Figures \ref{fig_paraFIR}, \ref{fig_paraMIR} and \ref{fig_paraNIR} show the $p$ and $T$ dependence of the peak frequency $\omega_0$ (a,b), the width $\gamma$ (c,d) and the oscillator strength $S$ (e,f) of the FIR, MIR and NIR h.o. Table \ref{Tab} summarizes the same fit parameters for the two visible (VIS-1/2) h.o.'s, which turn out to be totally $p$- and $T$-independent.

We shall just point out a few characteristic features for each h.o. For the FIR h.o. (Fig. \ref{fig_paraFIR}), $\omega_0$ is almost $p$- and $T$-independent and is isotropic between both axes. The anisotropy of its width develops because of the pronounced narrowing at $T << T_{s,N}$ along the $a$ axis for $p \sim$ 0.8 bar and upon releasing $p$ back to zero. The h.o.'s strength gets overall depleted with decreasing $T$ below 160 K for both axes. The anisotropy of this parameter is mainly due to its subsequent strong enhancement at $T << T_{s,N}$ along the $b$ axis for $p \ge$ 0.6 bar and upon releasing $p$ back to zero.

The peak frequency of the MIR h.o. (Fig. \ref{fig_paraMIR}) is enhanced when crossing $T_{s,N}$ and is anisotropic below $T_{s,N}$, since the frequency up-shift along the $b$ axis is pretty remarkable and stronger than along the $a$ axis. On the other hand for $p \sim$ 0.8 bar and upon releasing $p$ back to zero at $T << T_{s,N}$, there is a narrowing of the MIR h.o. along the $a$ axis. The h.o.'s strength gets predominately enhanced in the orthorhombic phase and particularly along the $b$ axis for $p \ge$ 0.4 bar as well as upon releasing $p$ back to zero at $T << T_{s,N}$.

As far as the NIR h.o. is concerned (Fig. \ref{fig_paraNIR}), its peak frequency shifts in an isotropic fashion to lower values at $T \le T_{s,N}$ for all $p$, while its width remains constant for both crystallographic directions. The h.o.'s strength is slightly enhanced for $p \le$ 0.2 bar along the $a$ axis at $T \le T_{s,N}$, prior recovering a $T$-independent value upon applying $p$, as well as releasing it. Along the $b$ axis, the strength in the orthorhombic phase is enhanced for all $p$, somehow in a stronger manner for 50 $\le T \le$ 70 K close to saturation (i.e., at 0.8 and released 0.5 bar).

The $p$ and $T$ dependence of the fit parameters, particularly of the h.o. strength, has relevant implications on the overall spectral weight reshuffling and redistribution, which will be addressed in a forthcoming review paper.

Finally, we shall briefly comment on the IR-active phonon mode (OP in Fig. 2(e)), which is only detected along the $b$ axis at $T \le T_{s,N}$, while it disappears above $T_{s,N}$ because of the screening due to the metallic components of $\sigma_1(\omega)$. At $T \le T_{s,N}$, the phonon fit parameters are almost $p$- and $T$-independent so that the peak frequency is at about 260 cm$^{-1}$, its width lies between 3 to 10 cm$^{-1}$ and the strength amounts to approximately 100 cm$^{-1}$. These latter values are very much consistent with the experimental finding of Ref.  \onlinecite{schafgans}. The OP contribution to the fit procedure is however quite negligible and does not affect its quality.

\end{document}